\documentstyle[11pt,newpasp,twoside,psfig,epsf]{article}
\def\ltsima{$\; \buildrel < \over \sim \;$}
\def\simlt{\lower.5ex\hbox{\ltsima}}
\def\gtsima{$\; \buildrel > \over \sim \;$}
\def\simgt{\lower.5ex\hbox{\gtsima}}

\def\edcomment#1{\iffalse\marginpar{\raggedright\sl#1\/}\else\relax\fi}
\reversemarginpar

\begin{document}
\title{The chemical enrichment of the ICM with BeppoSAX}
\author{Sabrina De Grandi}
\affil{Osservatorio Astronomico di Brera, via E. Bianchi 46, 23807 Merate (LC),
Italy}
\author{Silvano Molendi}
\affil{Istituto di Fisica Cosmica ``G. Occhialini'', via Bassini 15, 20100 
Milano, Italy}

\begin{abstract}
We review observations on the chemical enrichment of the intracluster
medium (ICM) performed using BeppoSAX MECS data.
The picture emerging is that non-cooling flow clusters have flat
metallicity profiles, whereas a strong enhancement in the abundance is
found in the central regions of the cooling flow clusters.
All the non-cooling flow clusters present evidence of recent merger
activity suggesting that the merger events redistributes efficiently
the metal content of the ICM.
The observed abundance excess in the central regions of cooling flow
clusters is probably due to metals ejected from the cD galaxy located
in the cluster core.
Cooling flow cluster have also enhanced Nickel abundances in their
cores with respect to the non cooling flow clusters. 
\end{abstract}

\section{Introduction}
The X-ray emission in clusters of galaxies originates from the hot gas
permeating the cluster potential well. The continuum emission is
dominated by thermal bremsstrahlung, which is proportional to the
square of the gas density times a function of the gas temperature (i.e.
the cooling function). From the shape and the normalization of the
spectrum we derive the gas temperature and density. At the high
temperatures measured in clusters the ICM is
highly ionized (H and He are completely ionized).
A measure of the equivalent width of a spectral line is a direct
measure of the relative abundance of a given element. This comes
from the fact that both the continuum and line emissions are two-body
processes with the continuum emissivity proportional to the electron density
times the proton density, and the line emissivity
proportional to the electron density times the density of a given
element. From the definition of equivalent width
it is easily derived that this quantity is proportional to the ratio
between the ion and proton densities.

In the 2-10 keV band Iron is the most easily measurable element 
in the X-ray spectrum. This is true for several reasons:
the atomic physics of K-shell lines is well known; Fe is an abundant
element in nature; the Fe-K$\alpha$ line is well isolated in the
spectrum and it is located in a region of the spectrum where the
spectral resolution of the X-ray detectors is usually good.  Other
elements measurable in the X-rays are K-shell lines of C,
N, O, Ne, Mg, Si, S, Ar, Ca and Ni, and L-shell lines of Fe and Ni.

It is well known that the global Fe abundance in clusters is about a
third of the solar value. The evolution of metal abundances of the ICM
in clusters has been investigated by various authors and a common
result is the lack of evolution within redshift about 0.3-0.4
(e.g. Allen \& Fabian 1998). This is also confirmed by BeppoSAX
measures on distant clusters (Della Ceca et al. 2000; Ettori, Allen \&
Fabian 2001).

While the origin of the metals observed in the ICM is clear (SNs make
metals), less clear is the transfer mechanism of these metals from
stars to the ICM. The main mechanisms that have been proposed for the
metal enrichment in clusters are: enrichment of gas during the
formation of the proto-cluster (e.g. Gnedin 1998, Kauffmann \& Charlot
1998) ; ram pressure stripping of metal enriched gas from cluster
galaxies (e.g. Gunn \& Gott 1972, Toniazzo \& Schindler 2001) ;
stellar winds AGN- or SN-induced in Early-type galaxies
(e.g. Matteucci \& Vettolani 1988, Renzini 1997). As we shall see
below clues to metal enrichment mechanisms can be derived from
spatially resolved analysis of metal abundances.

An interesting result is that found by Allen \& Fabian (1998) for a
sample of cooling flow (CF) and non-cooling-flow (non-CF) clusters,
where these authors found that CF clusters have global metallicities
about 1.8 times higher that that of non-CF clusters. This difference
could be explained by assuming that CF clusters have abundances excess
in their cores.
Understanding if the metal production in the ICM is segregated or if
it is constant with the cluster radius is important because it leads
to a more precise estimate of the metal amount in the ICM (e.g. the
total iron mass) and gives clues to the transport mechanism(s) from
galaxies to the ICM. Moreover, the comparison between the spatial
distribution of metals and the optical light gives information on the
galaxies which have contributed to the enrichment of the ICM.

\begin{figure}
\centerline{\psfig{figure=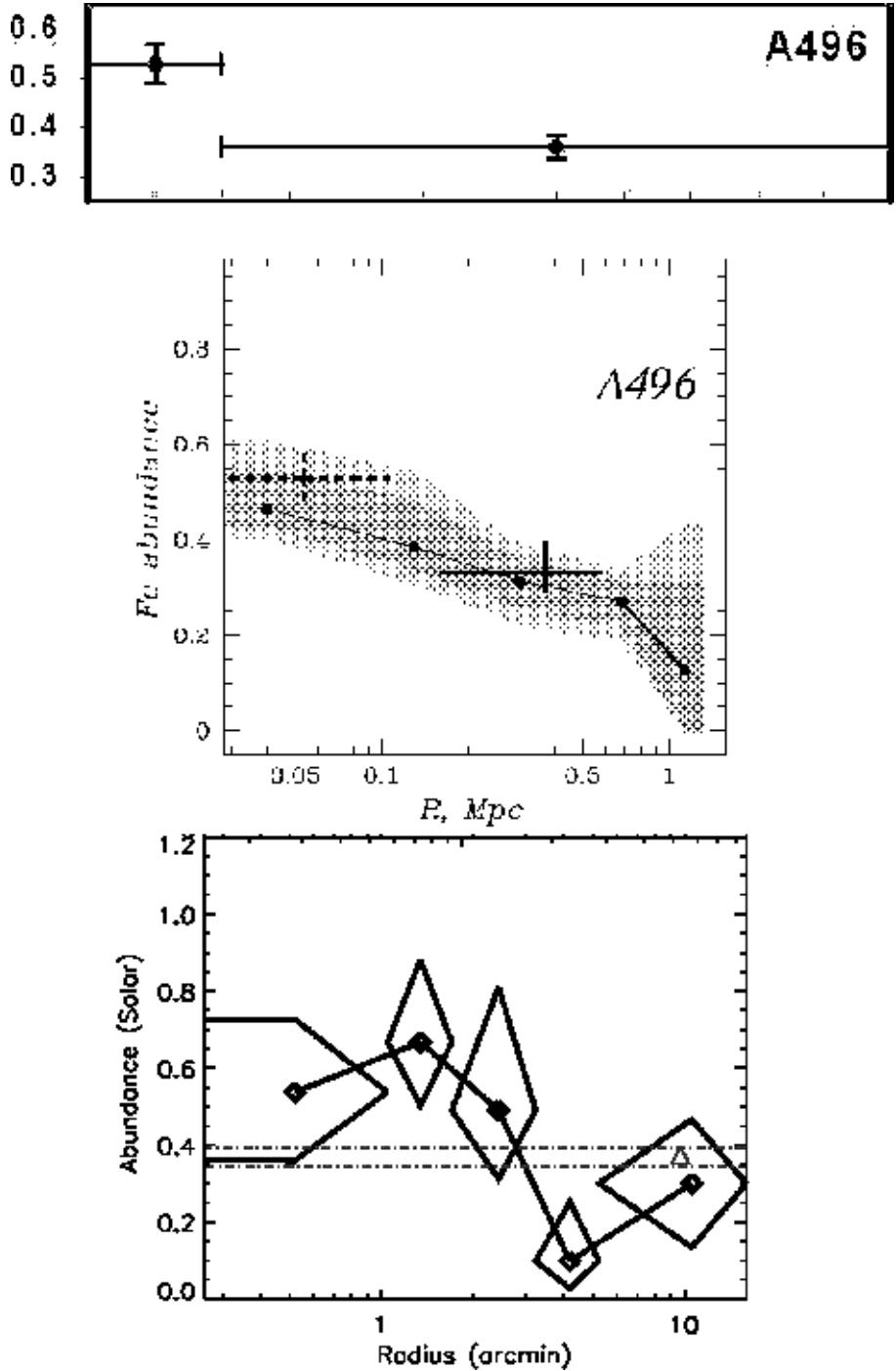,width=12cm,angle=0,clip=}}
\caption{Metallicity profile as a function of the radius in arcmin 
for cluster Abell 496 as observed with ASCA GIS. Top: Dupke \& White
(2000), the first bin ranges from 0 to 2 arcmin, the second bin from 2
to 12 arcmin. Middle: Finoguenov et al. (2000). Bottom: White (2000).}
\end{figure}

\section{Spatially resolved abundance measures in clusters observed 
with ASCA and BeppoSAX}

ASCA and BeppoSAX have been decisive missions to explore spatially
resolved abundance measurements in clusters. Various works have
identified abundance gradients in a few cooling flow clusters, e.g.
Ikebe et al. (1997) on Hydra A, Molendi et al. (1998) on Perseus
cluster, Dupke and White (2000) on A496.

Metal abundance derived from ASCA data for samples of rich clusters
are reported in Kikuchi et al. (1999), White (2000), Fukazawa et
al. (1998), Fukazawa et al. (2000), Dupke and White (2000),
Finoguenov, David and Ponman (2000). A general result from these works
is the evidence of abundance gradients in several clusters, with an
indication that abundance gradients are common in CF clusters, however
the shape of these gradients is poorly determined.  In fact the
quality of the ASCA data does not allow a detailed investigation of
the abundance gradients. A typical example is the case of A496 where
abundance profile has been derived from ASCA data by three different
working groups. In the first case Dupke \& White (2000) derived an
abundance gradient (see Figure 1 top panel) dividing the cluster
emission into two bins only, one from 0 to 2 arcmin and the second one
from 2 to 12 arcmin. In the second case Finoguenov et al. (2000) by
increasing the number of bins (at the price of increasing the errors
in the abundance measurements) up to four within 13 arcmin (Figure 1
middle panel) find evidences of for a smoothly declining gradient. In
the last case, similarly to Finoguenov et al, White (2000) increased
the numbers of bins up to 5 within 12 arcmin finding again an evidence
for a smoothly declining gradient with large errors (see Figure 1
bottom panel). A comparison of all these three cases shows a poor
determination of the real shape of the abundance gradient in A496. In
Figure 2 we show the abundance profile as observed with BeppoSAX, the
data are able to constrain the shape of the gradient revealing that
the abundance excess is concentrated in the cluster core.

\begin{figure}
\centerline{\psfig{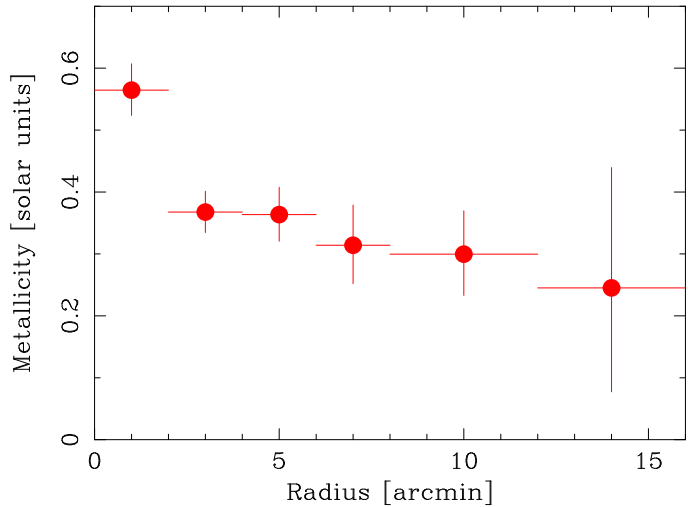}}
\caption{Metallicity profile as a function of the radius in arcmin 
for cluster Abell 496 as observed with BeppoSAX MECS.}
\end{figure} 

\begin{figure}
\centerline{\psfig{figure=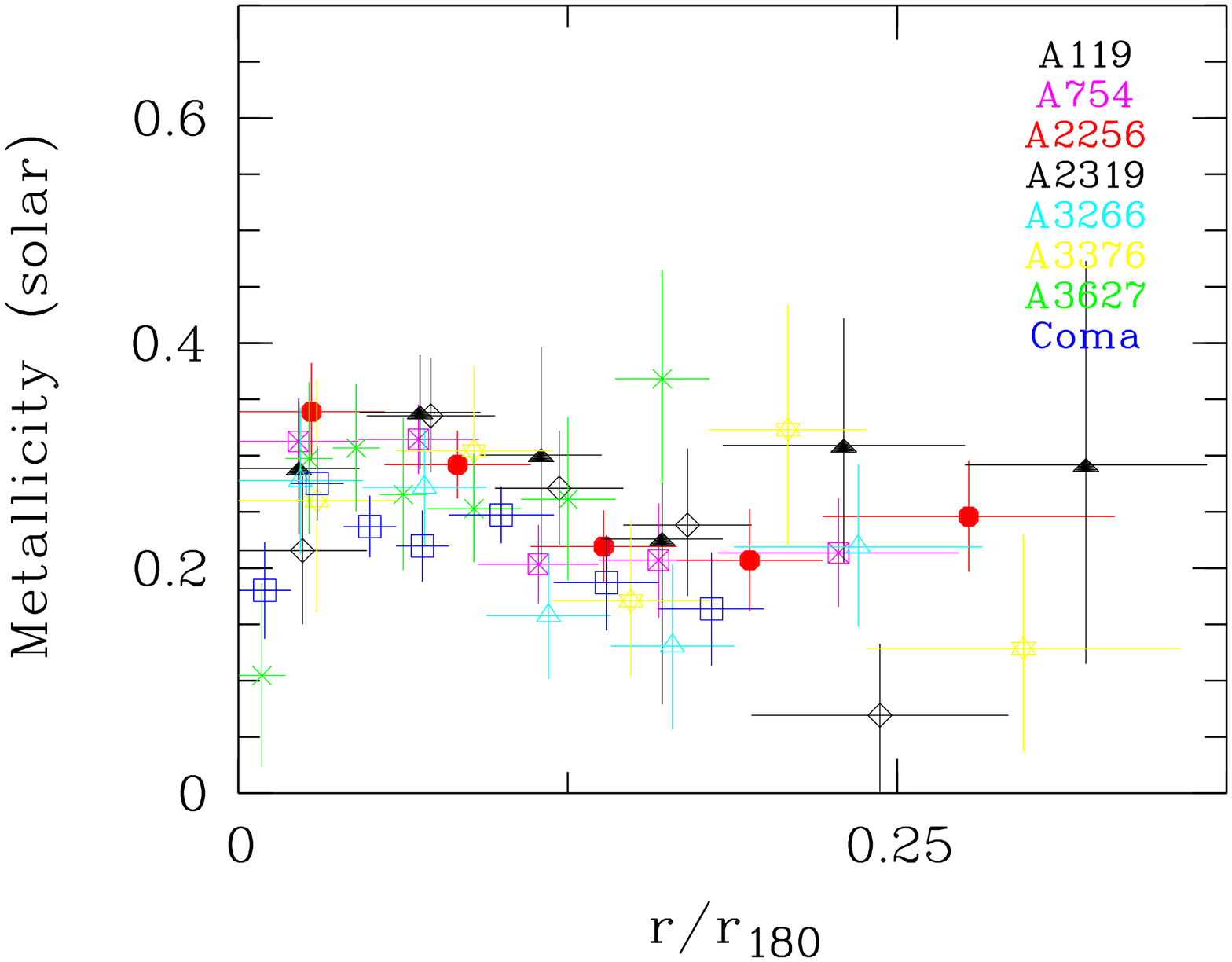,width=13cm,angle=-90,clip=}}
\centerline{\psfig{figure=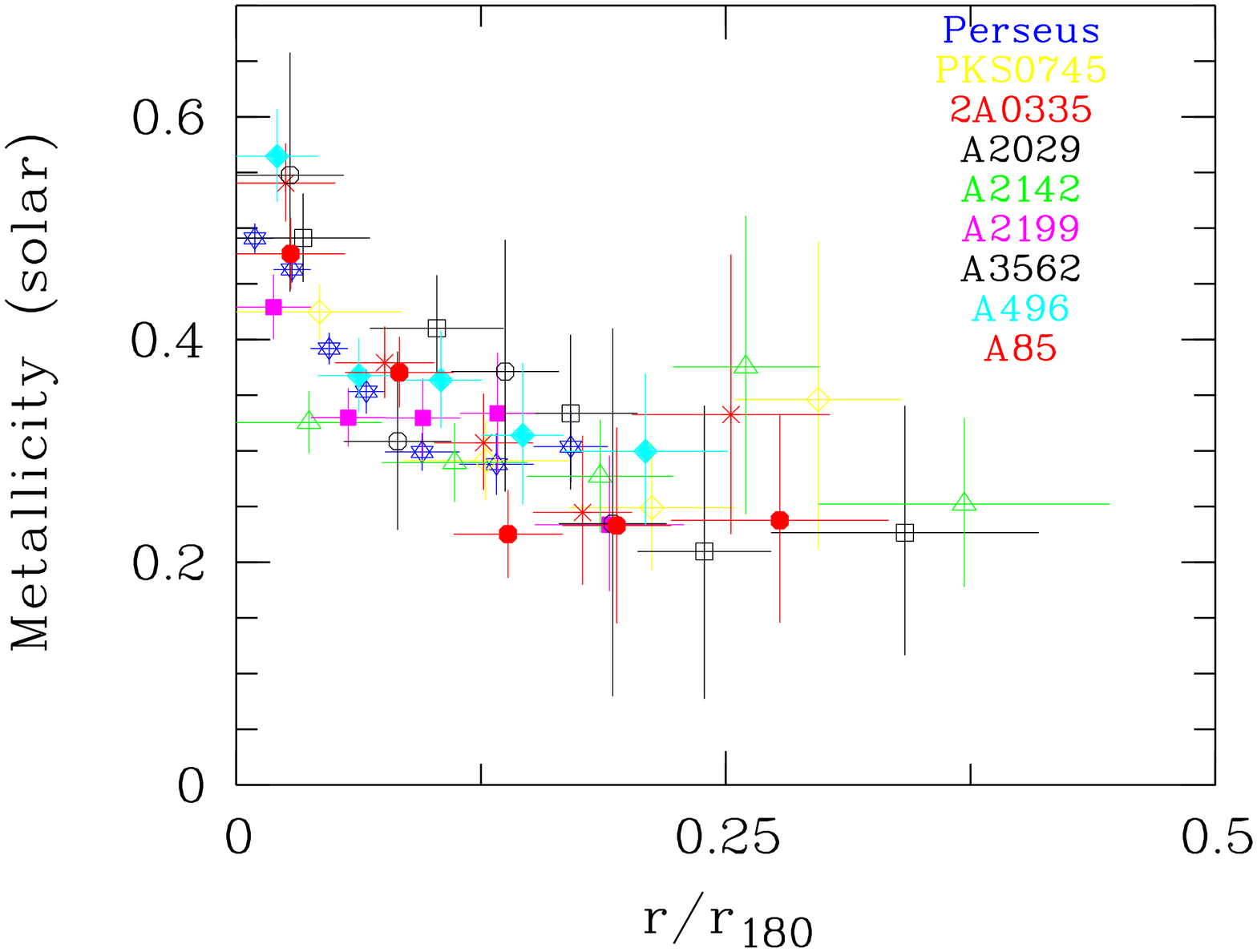,width=13cm,angle=-90,clip=}}
\caption{Metallicity profiles (projected) for the non-CF (top)
and CF (bottom) clusters, plotted against radii in units of
$r_{180}$.}
\end{figure} 

The difference between the ASCA and BeppoSAX metallicity profiles is
due to the differences in the PSFs. The ASCA PSF is broad (HPR $\sim
2$ arcmin), strongly energy dependent and non-radially symmetric,
therefore the analysis of extended sources required complicated
correction procedures (e.g. Markevitch et al. 1998, White \& Buote
2000). On the other hand, BeppoSAX PSF is sharper (HPR $\sim 1$
arcmin), is almost energy independent (D'Acri et al. 1997) and
radially symmetric, therefore the analysis of extended sources is
relatively straightforward (e.g. De Grandi \& Molendi 2001). Moreover
BeppoSAX exposure times are typically about three times larger than
ASCA exposure times. From all these considerations it follows that
BeppoSAX MECS data are better suited than ASCA to investigate
abundance profiles for galaxy clusters.

\begin{figure}
\centerline{\psfig{figure=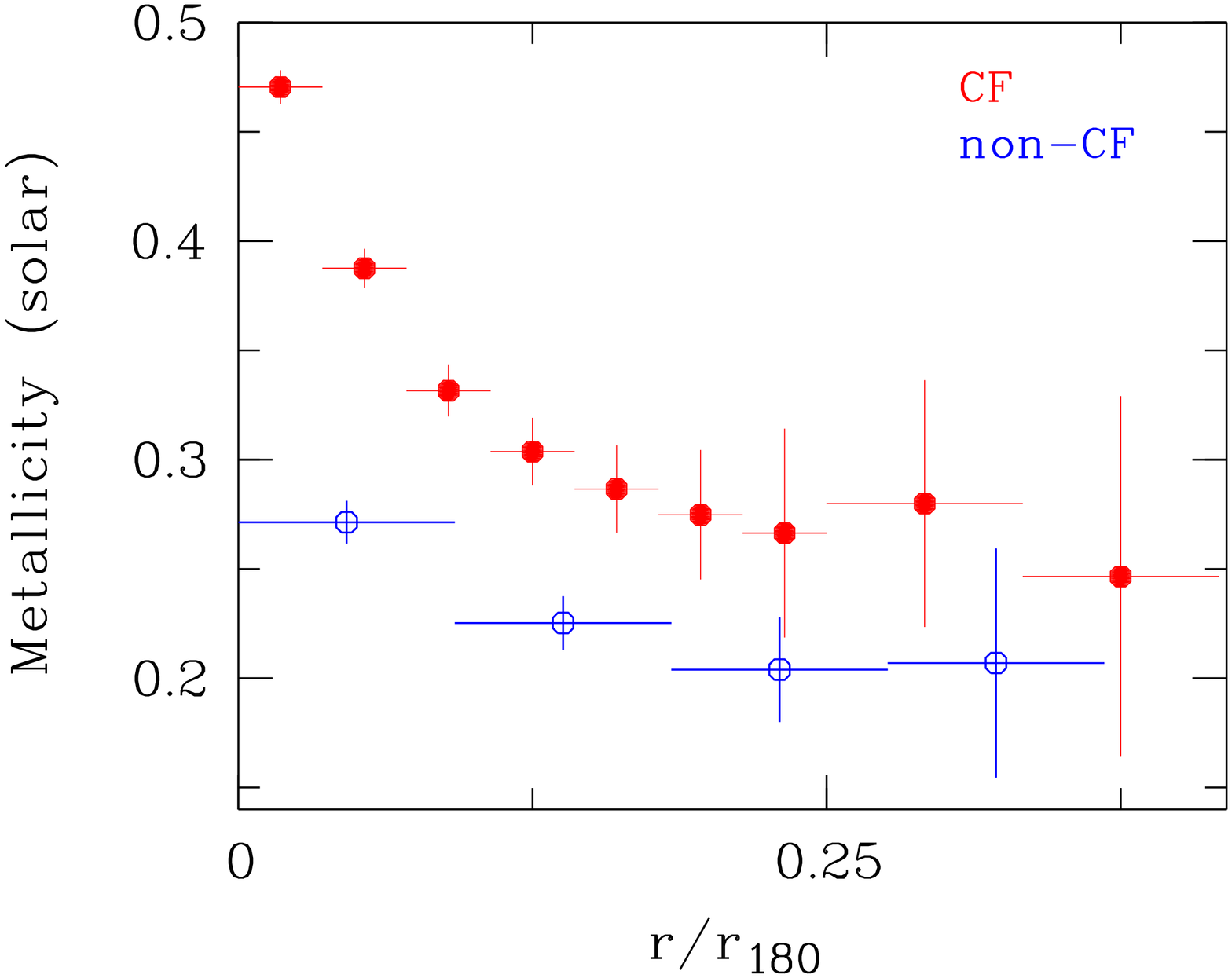,width=13cm,angle=-90,clip=}}
\caption{Mean metallicity profiles for the CF (filled circles) and
non-CF clusters (open circles), plotted against radii in units of
$r_{180}$.}
\end{figure} 

Up to date there are two works based on BeppoSAX MECS samples which are
systematically searching for abundance gradients in clusters. The first
is an analysis of 12 clusters performed by Irwin \& Bregman (2001).
This analysis is limited to the 9 innermost arcmin of the cluster
emission (i.e. radii $\simlt 20\%$ of the virial radius) and do
not explore systematically the difference between CF and non-CF
clusters.  Irwin \& Bregman (2001) find a general evidence for a
negative abundance gradient in most of the clusters.

The other work is that of De Grandi \& Molendi (2001) on a sample of
17 clusters analyzed considering the whole field of view of the MECS
(which corresponds to radii $\simlt 50\%$ of the virial radius).
The projected abundance profiles of the non-CF systems (Figure 3
top panel), are consistent with being constant with the radius.  On
the contrary the metallicity profiles of the CF clusters (Figure 3
bottom panel) is completely different showing a clear evidence of an
abundance gradient declining with the radius in most of the systems.

In Figure 4 we compare the mean error-weighted abundance profile for
CF and non-CF clusters.  The metal abundances of the CF clusters are
larger than 0.4 of the solar value in the central regions and decrease
rapidly to values similar to those of the non-CF clusters at radii
$\simgt 0.25~ \rm {r_{180}}$.  
The profile for non-CF clusters is much flatter, a fit with a constant
to all non-CF abundance measurements is statistically acceptable.
However, a small gradient is present in the data (significant at more
than $99.5\%$ level on the basis of an F-test).
The comparison of the abundance profile for CF and non-CF clusters
supports the scenario where major merger events disrupt the central
regions of clusters thereby re-mixing the gas within these regions and
therefore destroying pre-existing abundance gradients.  The modest
gradient observed in non-CF clusters is quite likely the relic of a
much stronger gradient which has not been completely wiped out by
merger events.

\subsection{Implication for the enrichment mechanism of the ICM in 
CF cluster cores}

If each merger events redistributes efficiently the metals within the
ICM then the metal excess we see in the core of CF clusters should be
directly related to the enrichment processes which have occurred in
the cluster core since the last major merger. Thus, just as the global
metallicity of clusters is an indicator of the global star formation
history within the whole cluster, the abundance excess we see in the
core of CF clusters is an indicator of the star formation history in
the core of the cluster since the last major merger.  In the light of
the above statement, we have tried to test whether the metal abundance
excess we see is due to metals expelled from early type galaxies
located in the core of the cluster. More specifically we have computed
the metal abundance excess profile expected when the metal excess
distribution traces the light distribution of early type galaxies,
included the cD galaxy (details are given in De Grandi \& Molendi
2001).
We have performed this computation for the 4 cooling flow objects
where the metal abundance profile is best measured and optical data is
available, namely: A85, A496, A2029 and Perseus (see Figure 5). 

\begin{figure}
\centerline{\psfig{figure=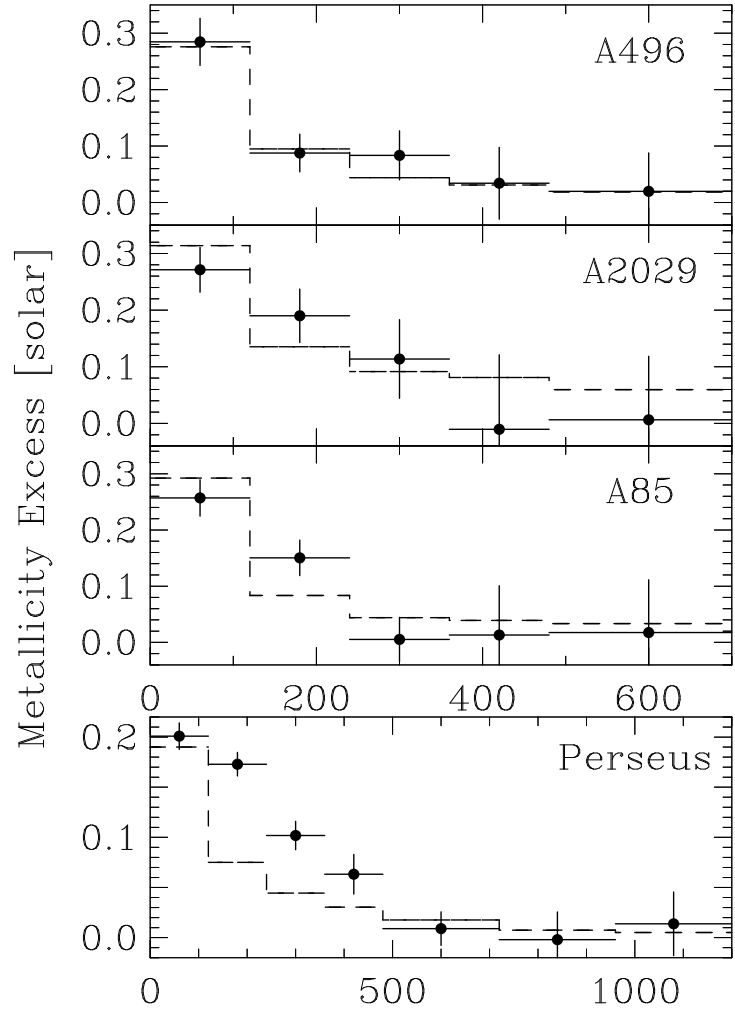,width=20cm,angle=-90,clip=}}
\caption{Predicted (dashed lines) versus measured (solid circles) metallicity
excess profiles as a function of radius for the cooling flow clusters
A496, A2029, A85 and Perseus.}
\end{figure} 

\begin{figure}
\centerline{\psfig{figure=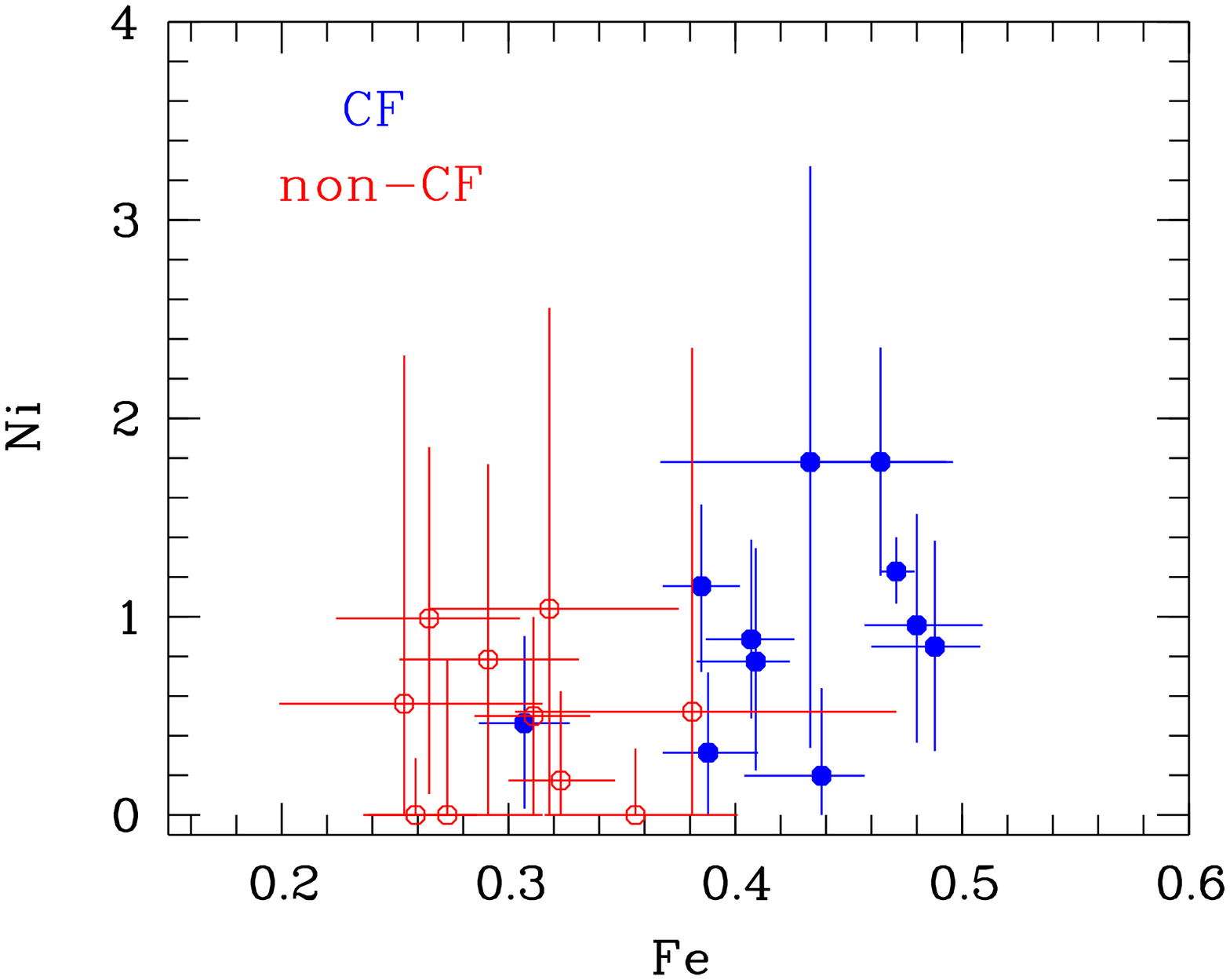,width=13cm,angle=-90,clip=}}
\caption{Nickel vs. Iron abundances for non-CF (open circles) and
CF (filled circles) clusters.}
\end{figure} 

For A496 and A2029 the predicted projected metal abundance excess can
be reconciled with the observed one, while for A85 the predicted
projected metal abundance excess appears to be slightly more centrally
concentrated than the observed one. The more interesting case is that
of Perseus cluster, where the two profiles are substantially different.

We find that the abundance excess in the expected profiles is
completely due to the cD galaxy. Fukazawa et al. (2000) computed that
the amount of metal excess at the cluster center can be provided by the
cD alone. Therefore we conclude that we are probably just observing
the accumulation of metal ejection from the cD galaxy into the ICM,
and suggest that a possible way of reconciling the observed and
predicted abundance excess profile of Perseus is to assume that metals
ejected from the central cD have drifted away by about 50 kpc.

\section{BeppoSAX Nickel measurements}

For an updated sample of 22 clusters (11 with CF and 10 without CF) we
have measured Nickel abundances in the innermost regions of the
clusters, i.e. for radii smaller than 200-400 kpc.  The results are
reported in Figure 6. This figure shows the Fe segregation
between CF and non-CF clusters already discussed in the previous
sections, as well as a difference for the Ni abundances.

The mean Fe abundance values for CF and non-CF clusters are, in solar
units, $0.45\pm0.01$ and $0.30\pm0.01$, whereas the mean Ni abundances
are $1.10\pm0.12$ and $0.16\pm0.17$, respectively. We find a $\sim
5\sigma$ evidence for a Ni excess in the central regions of the CF
clusters with respect to the central regions of non-CF clusters. This
Nickel excess is probably associated to the cD galaxy as the Iron
excess, indeed both elements are mostly produced by SN type Ia which
are dominating the enrichment of the cluster central regions. The
Nickel-to-Iron abundance ratio produced by the cD galaxy, defined as
$[Ni/Fe]_{cD} = ([Ni]_{CF} - [Ni]_{non-CF}) / ([Fe]_{CF} -
[Fe]_{non-CF})$, is $6.3\pm1.4$, normalized to the solar ratio.  

\section{References}
Allen, S. W., \& Fabian, A., C. 1998, MNRAS, 297, L63 (AF98) \\
Anders, E., \& Grevesse, N. 1989, Geochim. Cosmoshim. Acta, 53, 197 \\
D'Acri, F., De Grandi, S., \& Molendi S. 1998, Nuclear Physics,
69/1-3, 581 (astro-ph/9802070) \\
De Grandi, S. \& Molendi, S. 2001, ApJ, 551, 153 \\
Della Ceca, R., Scaramella, R., Gioia, I. M., Rosati, P., Fiore, F.,
Squires, G. 2000, A\&A, 353, 498 \\
Dupke, R. A. \& White R. 2000, ApJ, 528, 139 \\
Ettori, S., Allen, S. W., Fabian, A. C. 2001, MNRAS, 322, 187 \\
Finoguenov, A., David, L. P. \& Ponman, T. J.  2000, ApJ, 544, 188 \\
Fukazawa, Y., Makishima, K., Tamura, T., Ezawa, H., Xu, H., Ikebe,
Y., Kikuchi, K. \& Ohashi, T. 1998, PASJ, 50, 187 \\
Fukazawa, Y., Makishima, K., Tamura, T., Nakazawa, K., Ezawa, H.,
Gnedin, N. Y. 1998, MNRAS, 294, 407 \\
Gunn, J. E. \& Gott J. R. 1972, ApJ, 176, 1 \\
Ikebe, Y., Kikuchi, K., Ohashi, T. 2000, MNRAS, 313, 21 \\
Ikebe, Y. et al. 1997, ApJ, 481, 660\\
Irwin, J. A. \& Bregman J. N. 2001, ApJ, 546, 150\\
Kauffmann, G. \& Charlot, S.  1998, MNRAS, 294, 705 \\
Kikuchi, K., Furusho, T., Ezawa, H., Yamasaki, N. Y., Ohashi, T.,
Fukazawa, Y.; Ikebe, Y. 1999, PASJ, 51, 301 \\
Markevitch, M., Forman, W. R., Sarazin, C. L. \& Vikhlinin, A. 1998,
 ApJ, 503, 77 \\
Matteucci, F. \& Vettolani, G.  1988, A\&A, 202, 21 \\
Molendi, S., Matt, G., Antonelli, L. A., Fiore, F., Fusco-Femiano, R.,
Kaastra, J., Maccarone, C. \& Perola, C. 1998, 499, 608 \\
Renzini, A. 1997, ApJ, 488, 35 \\
Toniazzo, T. \& Schindler, S. 2001, MNRAS in press (astro-ph/0102204) \\
White, D. A. 2000, MNRAS, 312, 663 \\
White, D. A. \& Buote, D. A., 2000, MNRAS, 312, 649 \\
\end{document}